\documentclass[letterpaper]{article}
\usepackage{aaai}
\usepackage{times}
\usepackage{helvet}
\usepackage{courier}
\usepackage{booktabs}
\IfFileExists{adjustbox.sty}{\usepackage{adjustbox}}{}
\usepackage{microtype}
\usepackage{amsmath}
\usepackage{amssymb}
\usepackage{amsfonts}
\usepackage{graphicx}
\usepackage{subcaption}
\usepackage{multirow}
\usepackage{float}
\usepackage{xcolor}
\usepackage{tikz}
\usetikzlibrary{arrows.meta,positioning,fit,backgrounds,calc}
\usepackage{caption}
\usepackage{placeins}
\usepackage{hyperref}

\hypersetup{
  colorlinks=true,
  linkcolor=[rgb]{0.10,0.20,0.75},
  citecolor=[rgb]{0.10,0.20,0.75},
  urlcolor=[rgb]{0.05,0.35,0.85},
  pdftitle={Towards Trustworthy Biological Alignment in TabPFN-Probed Pathology Foundation Models},
  pdfauthor={Ushashi Bhattacharjee, Alloy Das, Saria Hannan, Tirtho Roy, Koushik Howlader, Soumik Sarkar},
  pdfkeywords={pathology foundation models, spatial transcriptomics, in-context learning, trustworthy ML, distribution shift, shortcut learning}
}
\graphicspath{{figures/}{./}}
\newcommand{\covSamples}{240}
\newcommand{\covOrgans}{3}
\newcommand{\covTech}{4}
\newcommand{\ladTabPath}{0.303}\newcommand{\ladTabGene}{0.135}
\newcommand{\ladMlpPath}{0.161}

\newcommand{\ctrlMatchPath}{0.303}
\newcommand{\ctrlShufPath}{0.016}
\newcommand{\ctrlRandPath}{-0.006}
\newcommand{\ctrlWSecPath}{0.120}
\newcommand{\oodIID}{0.303}\newcommand{\oodOrgan}{0.114}\newcommand{\oodDrop}{62}
\newcommand{\perturbCleanPCC}{0.294}\newcommand{\perturbRetainGeoMin}{95}\newcommand{\perturbRetainMin}{72}\newcommand{\perturbAgreeMin}{0.84}

\title{Towards Trustworthy Biological Alignment in TabPFN-Probed Pathology Foundation Models}
\author{Ushashi Bhattacharjee$^{*}$, Alloy Das$^{*}$, Saria Hannan, Tirtho Roy,\\
Koushik Howlader, Soumik Sarkar$^{\dagger}$\\
{\normalsize\normalfont $^{*}$Equal contribution. \quad $^{\dagger}$Corresponding author.}}
\nocopyright

\begin{document}
\maketitle

\begin{abstract}
\begin{quote}
Histology and transcriptomic data provide complementary views of tissue biology through spatial morphology and molecular activity. However, pathology foundation models (PFMs) encode rich tissue morphology, but strong downstream performance does not necessarily indicate that their representations capture robust biological information. We present a training-free framework for auditing biological alignment in frozen PFMs using spatially paired histology and transcriptomics from HEST-1k, evaluated on \covSamples{} samples spanning \covOrgans{} organs. We use several frozen PFMs to get features from H\&E images and group gene expression into meaningful biological pathways. TabPFN serves as a pretrained probe to measure how well these molecular programs can be decoded without task-specific gradient updates. We test whether pathway predictions remain reliable across different tissue sections, patient groups, and tissue types, and whether the models rely on shortcuts or small image changes. We further track prediction stability under context resampling as a sensitivity diagnostic. The resulting multi-tissue audit characterizes which molecular pathways are robustly represented, tissue-specific, or shortcut-sensitive, providing a systematic approach for evaluating trustworthy biological alignment in pathology foundation models.

\end{quote}
\end{abstract}

\section{Introduction}
Spatial transcriptomics provides a molecular map of tissue by measuring thousands of genes and biological programs while retaining their spatial context. Pathology foundation models (PFMs) are promising for linking molecular measurements with tissue morphology because their pretrained representations capture complex patterns from H\&E images. Recent PFMs, including UNI, Virchow, Prov-GigaPath, and Phikon-v2, have demonstrated strong performance across computational pathology tasks. However, successful downstream prediction does not necessarily imply that these representations capture genuine molecular biology. Apparent morphology–molecular associations can instead reflect section identity, staining variation, spatial autocorrelation, or other dataset-specific shortcuts. For reliable biological interpretation, strong molecular decodability alone should not be treated as evidence of meaningful biology. A trustworthy evaluation should therefore answer two questions: what molecular programs are encoded in frozen pathology representations, and which remain reliable under distribution shifts and shortcut controls? 

Existing methods address these two questions only partially. Histology-to-expression methods such as STNet~\cite{he2020stnet}, BLEEP~\cite{xie2023bleep}, and mclSTExp~\cite{min2024mclstexp} predict molecular states from tissue morphology, while multimodal approaches such as TANGLE~\cite{jaume2024tangle}, SPADE~\cite{spade2024}, SEAL~\cite{seal2024}, and PathLUPI~\cite{pathlupi2024} use transcriptomic supervision to improve morphology–molecular alignment. However, these approaches train task-specific components, making it difficult to determine whether the decoded biology was already present in the frozen representation or learned through downstream optimization. Recent studies also show that pathology foundation models can learn shortcuts from section, site, scanner, or batch information~\cite{dejong2025medical,komen2026robust,geirhos2020shortcut}. However, these shortcuts are usually studied separately from molecular prediction. As a result, existing methods do not jointly examine what molecular information is encoded in frozen PFMs and whether this information remains reliable under distribution shifts and shortcut controls.

\begin{figure*}[t]
    \centering
    \includegraphics[width=0.94\textwidth]{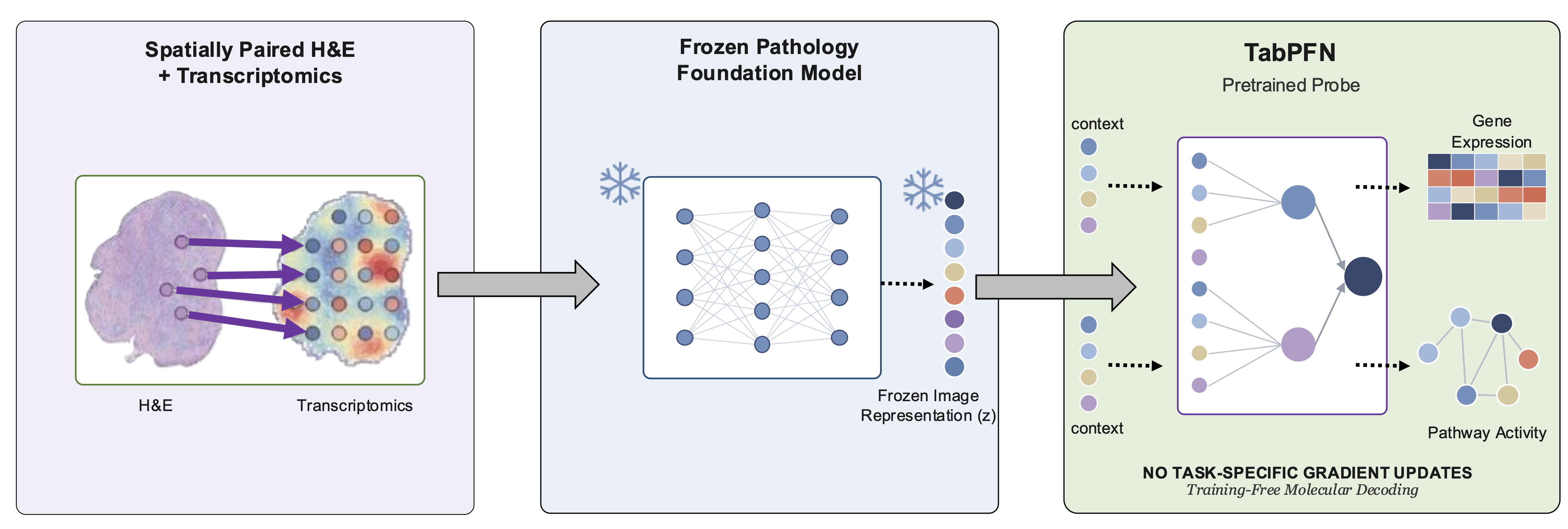}
    \caption{\textbf{Training-free morpho--molecular audit.} Spatially paired H\&E and
    transcriptomics from HEST-1k are encoded by a \emph{frozen} pathology foundation model
    (UNI, Phikon, Phikon-v2, Lunit, or CTransPath); a \emph{frozen} TabPFN probe then decodes
    gene expression and pathway activity in-context. No parameters are updated, so decodability
    reflects the representation and the underlying biology rather than a trained prediction head.}
    \label{fig:methodology}
\end{figure*}

To address this gap, we present a training-free framework for auditing biological alignment in frozen pathology foundation models. Rather than training a new predictor, we ask what molecular information is already present in their representations and whether that information can be trusted. As shown in Figure~\ref{fig:methodology}, we extract representations from H\&E images using frozen PFMs and pair them with spatially matched gene expression from HEST-1k. Gene expression is further organized into biologically meaningful pathway activities, and TabPFN is used as a pretrained probe to decode genes and pathways from the frozen representations without task-specific gradient updates. Using the same frozen probing framework, we then evaluate whether the decoded signal survives negative controls, distribution shifts, image perturbations, and shortcut tests. In this way, TabPFN serves as a standardized audit probe rather than a new downstream prediction model. 

We operationalize these two questions through four concrete audit questions: whether the training-free audit recovers molecular information from frozen representations; how molecular decodability varies across gene and pathway targets and tissues; whether the recovered signal remains reliable under distribution shift while separating biological alignment from shortcuts; and when the resulting molecular readout is trustworthy under matched, shuffled, random-feature, and stability controls. Across \covSamples{} samples spanning \covOrgans{} organs (Breast, Skin, Brain), we find that molecular decodability is strongly tissue-dependent and that pathway activity is more consistently recoverable than individual gene expression. We further find that strong decodability does not always translate into trustworthy alignment: performance degrades under distribution shift, while section identity can remain strongly encoded in the representations. These results show why molecular prediction alone is insufficient for evaluating biological alignment in pathology foundation models.

In summary, our work makes the following contributions:

\begin{itemize}
    \item We introduce a training-free framework for auditing biological alignment in frozen pathology encoders using spatially paired histology and transcriptomics, with TabPFN serving as a standardized pretrained probe rather than a task-specific predictor.

    \item We define an audit protocol that couples gene- and pathway-level decodability with negative controls (label-shuffle, random-feature), distribution-shift evaluation, benign-perturbation stability, and a within-section shortcut test, all applied to the same frozen pipeline.

    \item We show that molecular decodability and trustworthiness can diverge: frozen pathology representations carry biologically patterned molecular signal that survives our controls, yet part of their apparent alignment is sensitive to distribution shift or driven by section-level shortcuts (section identity is itself linearly decodable up to $214\times$ chance).
\end{itemize}

\paragraph{Related work.}
\textbf{Pathology foundation models.} General-purpose pathology encoders such as
UNI \cite{chen2024uni}, CONCH \cite{lu2024conch}, Virchow
\cite{vorontsov2024virchow}, Prov-GigaPath \cite{xu2024gigapath}, and Phikon-v2
\cite{filiot2024phikonv2}
learn transferable visual representations from large
histology collections. Their papers evaluate these representations across diverse
downstream tasks, including frozen or linear-probe settings. Such transfer results
are evidence of useful representations, but they do not by themselves establish
whether an embedding carries molecular biology or dataset-specific technical cues.

\textbf{Histology-to-molecular prediction.} STNet
\cite{he2020stnet}, BLEEP \cite{xie2023bleep}, mclSTExp \cite{min2024mclstexp}, and
TANGLE \cite{jaume2024tangle} demonstrate that transcriptomic information can be
associated with tissue morphology. Related alignment and molecularly supervised
methods---including TANGLE, UMPIRE, THREADS, MINT, SPADE, SEAL, and
PathLUPI~\cite{jaume2024tangle,umpire2024,threads2025,mint2026,spade2024,seal2024,pathlupi2024}---
inject transcriptomic or genomic information during pretraining, fine-tuning, or
fusion. They are important complementary models and useful reference points for
robustness, but they do not answer the same diagnostic question: their performance
reflects both the initial visual representation and learned molecular supervision,
whereas our audit isolates information accessible before task-specific updates. We
therefore discuss, but do not treat, these trained systems as like-for-like empirical
baselines for the frozen audit.

\textbf{Frozen probing and in-context prediction.} TabPFN is a tabular foundation
model that conditions on labeled examples at inference time
\cite{hollmann2023tabpfn,hollmann2025tabpfnv2}; it is not itself a pathology-specific
alignment method. Recent work has begun to study frozen visual features with
in-context or frozen prediction heads, including
TIME \cite{time2024} and Images as Tables \cite{walter2026images}.
These studies motivate using a fixed tabular model as a probe, but they do not by
themselves establish a standardized audit of morpho--molecular information in
pathology embeddings. We use TabPFN in that narrower, explicitly diagnostic role.

\textbf{Trustworthiness and shortcut auditing.} Shortcut-learning work shows that
predictive performance can arise from unintended correlations rather than
task-relevant information \cite{geirhos2020shortcut,degrave2021ai}. In pathology,
site, scanner, center, and section effects are plausible sources of such cues. This
motivates our matched, shuffled, random-feature, distribution-shift, and
within-section controls. Our contribution is a combination claim: we connect
general pathology foundation models, molecularly informed histology methods, and
robustness auditing in one training-free evaluation, rather than claiming novelty
for any individual component. Mechanistic approaches based on sparse autoencoders
decompose PFM embeddings into human-interpretable biological concepts
\cite{le2024sae}; these diagnose \emph{what} an embedding represents and are
complementary to our tests of whether a molecular readout survives controls and shift.

\section{Method}
\subsection{Preliminaries}
Let $\mathcal{D}=\{(x_i,y_i,s_i,g_i)\}_{i=1}^{N}$ be a collection of H\&E patches
$x_i$, molecular targets $y_i\in\mathbb{R}^{D}$ (gene expression or pathway
activity), section identities $s_i$, and specimen/patient groups $g_i$. A frozen
encoder $E$ maps each patch to an embedding $z_i=E(x_i)\in\mathbb{R}^{d}$ ($d{=}1024$
for UNI), and a frozen tabular model $F$ decodes a query embedding $z_q$ from a
labeled context $\mathcal{C}_K=\{(z_i,y_i)\}_{i=1}^{K}$ drawn from the pool:
\begin{equation}
\hat{y}_q=F\!\left(z_q\mid\mathcal{C}_K\right),\qquad
\text{no parameters of }E\text{ or }F\text{ change.}
\end{equation}
\emph{Decodability} is the mean per-target Pearson correlation between $\hat{y}$ and
$y$ over held-out queries,
$\rho=\tfrac{1}{D}\sum_{j}\mathrm{PCC}(\hat{y}^{(j)},y^{(j)})$. Because $E$ and $F$
are fixed, $\rho$ is a property of the \emph{representation and the biology}, not of
a trained head---this is what turns prediction into an audit.

Trustworthy alignment requires more than a large $\rho$. Figure~\ref{fig:audit}
summarizes the four audit operators, all applied to the same frozen pipeline.
\textbf{(1) Controls} replace the
context--target correspondence with a permutation $\pi$ ($y_i\!\mapsto\!y_{\pi(i)}$)
or replace $E$ with a random encoder; genuine signal must give
$\rho_{\text{matched}}\!\gg\!\rho_{\text{control}}\!\approx\!0$. \textbf{(2)
Distribution shift} draws $\mathcal{C}_K$ and $z_q$ from disjoint sections,
cohorts, or organs, measuring $\rho_{\text{OOD}}$ versus $\rho_{\text{IID}}$.
\textbf{(3) Perturbation stability} re-embeds queries under label-preserving
transforms $T$ (stain/color jitter, rotation, blur, JPEG) and reports
$\Delta\rho$. \textbf{(4) Shortcut probes} fit a linear classifier
$s\!\approx\!W z$ for section identity, and permute targets \emph{within} a section
($y_i\!\mapsto\!y_{\pi(i)}$ with $s_{\pi(i)}{=}s_i$); if $\rho$ survives, the score
reflects section-level structure rather than local morphology$\to$biology coupling.
A representation is trustworthy only if $\rho$ is high \emph{and} survives (1)--(4).

\subsection{Training-free audit framework}
\begin{figure}[t]\centering
\includegraphics[width=\columnwidth]{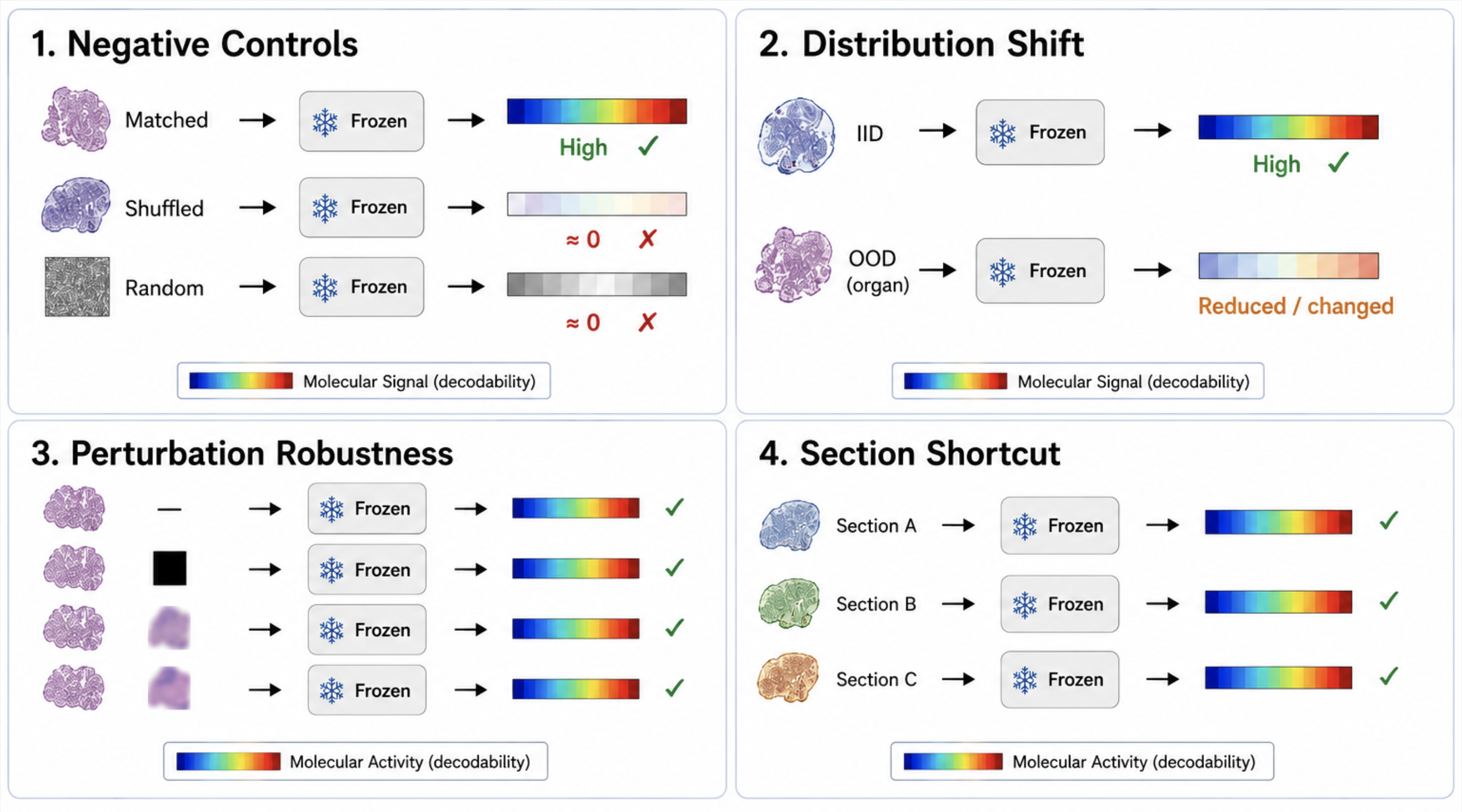}
\caption{\textbf{The four audit operators.} Applied to the same frozen pipeline:
\textbf{(1)} negative controls (matched vs.\ label-shuffled vs.\ random-representation),
\textbf{(2)} distribution shift (IID vs.\ organ-OOD), \textbf{(3)} benign image
perturbations, and \textbf{(4)} the section-identity shortcut. A representation is
trustworthy only if decodability is high \emph{and} survives all four.}
\label{fig:audit}
\end{figure}

\paragraph{Frozen probe.} $E$ is UNI (DINOv2 ViT-L/16
\cite{chen2024uni,oquab2024dinov2}); we use its frozen $1024$-d CLS embedding.
The probe is TabPFN's regression variant (\texttt{TabPFNRegressor}), used
\emph{continuously}: targets are \emph{not} discretized. Because TabPFN is a
single-output in-context model, a $D$-dimensional target is decoded as a bank of
$D$ independent one-dimensional regressors that share the same labeled context
$C_K=\{(z_i,y_i)\}$; \texttt{fit} merely stores $C_K$ and \texttt{predict} runs
one forward pass per target dimension. We use TabPFN's default regression ensemble
(exact library settings are in the released config); this is an implementation-level
ensemble setting, not feature selection or additional training. No weights change and
there is no per-target calibration beyond the target normalization above. Its only
inductive bias is TabPFN's synthetic-prior meta-training; the multi-encoder comparison
and the probe ladder (Ridge/kNN/MLP) test sensitivity to that bias, though absolute
scores and label-efficiency conclusions may remain probe-family dependent. Throughout,
\emph{training-free} means that neither the encoder $E$ nor the probe $F$ undergoes any
dataset-specific parameter update: TabPFN retains its synthetic-prior meta-training and
only conditions on the labeled context at inference.
\paragraph{Context construction.} For each query, the context $C_K$ of size
$K\!\in\!\{8,32,128,512\}$ is drawn from other specimens under one of four
strategies: \emph{random} (the default; uniform over the pooled training spots),
\emph{section-balanced} (equal quota per section, so no single section dominates),
\emph{diverse} (farthest-point sampling in cosine space), and per-query
\emph{nearest-neighbor}. Sampling uses a fixed seeded RNG; a query specimen never
contributes to its own context. We compare these strategies, and use
section-balanced contexts as an explicit check that the shortcut effect is not an
artifact of section-imbalanced sampling.
\paragraph{Targets and preprocessing.} Gene expression uses the top-1000 highly
variable genes per group; pathway activity is ssGSEA \cite{barbie2009ssgsea} over
Reactome \cite{gillespie2022reactome} $+$ MSigDB Hallmark
\cite{liberzon2015hallmark}. Raw counts are library-size normalized and
$\log$-transformed, with per-gene min--max scaling and $z$-scored pathway scores.
Because HEST mixes gene-symbol and Ensembl namespaces, we map all identifiers to a
common symbol space and form a per-(species,technology) panel from genes present
in $\geq$50\% of a group's samples. Because these panels differ across groups,
absolute PCC values are \emph{not} directly comparable between organs measured on
different technologies; we mitigate, but cannot eliminate, this confound by making every verdict a
\emph{within-group, within-organ} contrast (matched vs.\ its own within-section
control) rather than assuming absolute comparability across panels. Pathway (ssGSEA) scores
are additionally more panel-robust than single genes, as they aggregate over many
members of a set. We deliberately apply \emph{no} cross-study
batch correction to targets: correcting expression would partly determine the very
shortcut we audit, so we measure the batch/section shortcut directly.
\paragraph{Leakage-safe protocol.} Specimen-level grouping (patient where
available, else section); a query specimen never enters its own context. Five-fold
specimen-level cross-validation with three fixed seeds; we report mean per-target
Pearson correlation (PCC).
\paragraph{Baselines.} Beyond TabPFN we run mean, kNN (cosine, distance-weighted),
Ridge, and a gradient-trained MLP (one hidden layer of 256 units, ReLU, Adam
$10^{-3}$, weight decay $10^{-4}$, 150 epochs, standardized inputs, same context),
on identical folds and context, isolating head capacity from representation
quality.
\paragraph{Metrics and statistics.} Mean per-target PCC; per-organ/macro averages;
$K_{90}$. Unless noted, the unit of analysis is the per-target PCC (the $D$ pathway
or gene targets of an organ), aggregated within specimen-level folds and three seeds;
resampling for confidence intervals is done at the \emph{specimen} level (grouped
bootstrap), so spots from the same section are never treated as independent
replicates. Comparisons between methods are paired across matched targets, with
Wilcoxon and permutation tests and Holm correction over the organ$\times$target
family. TabPFN's advantage over MLP, Ridge, and label-shuffled is significant under
Holm-corrected permutation tests ($p<10^{-3}$).

\paragraph{Audit implementation.} The four operators (Fig.~\ref{fig:audit}) are
realized as follows. \emph{Negative controls} permute context labels or replace $E$
with random features. \emph{Distribution shift} holds out sections, cohorts, or whole
organs. Organ holdout is deliberately a stringent stress test of whether a learned
morphology--molecular mapping transports across biological domains; it does not assume
that molecular biology is invariant across organs. \emph{Benign perturbations} re-embed queries under stain/color jitter,
rotation/flip, blur, and JPEG at several severities, reporting the degradation
$\Delta\mathrm{PCC}=\mathrm{PCC}_{\text{clean}}-\mathrm{PCC}_{\text{perturbed}}$.
\emph{Shortcut probes} fit a linear section-identity classifier on $z$ and permute
targets \emph{within} a section; if performance survives, the score reflects
section-level structure rather than local morphology$\to$biology coupling. We
additionally track reseed stability---how much a prediction moves when only the
context draw changes---as a qualitative check that decoded state is a property of the
query rather than the sampled context.

\paragraph{Analysis cohort.} All analyses run on the same \covOrgans{} data-rich organs (Breast, Skin, Brain).
Each organ is assigned---\emph{post hoc}, from its measured matched-vs-control
contrast---to one of three regimes: \emph{genuine} (matched $\gg$ within-section
control), \emph{shortcut-suspect} (matched $\approx$ control), or \emph{weak}
(little signal). No organ is excluded; the regimes are an \emph{outcome} of the
audit, not a design choice.

\section{Experimental Setup}
We draw from HEST-1k \cite{jaume2024hest} (NeurIPS 2024 Datasets \& Benchmarks;
1{,}229 profiles, 26 organs, 2 species, $\sim$2.1M pairs, 153 cohorts). The audit is
inherently per-organ---each verdict compares an organ's matched decodability
against its \emph{own} within-section control---so a trustworthiness verdict is only
meaningful where an organ has enough sections to estimate that control. Many HEST
organs do not: lymph node and ovary contain just $5$ and $10$ profiles. We therefore
run the definitive audit on \textbf{\covOrgans{} data-rich organs} (Breast, Skin, and
Brain; each $n\!\geq\!37$, \covSamples{} samples total), selected solely because they
have enough sections per organ to estimate the within-section control.
Each organ contributes up to $80$ profiles and $800$ spots per section across all
\covTech{} spatial technologies. Regime membership is \emph{not} assigned a priori: each
organ is placed by its measured matched-vs-control contrast. Context sizes
$K\in\{8,32,128,512\}$; gene targets are the top-1000 HVGs and pathway targets the
full Reactome$+$Hallmark panel; five-fold specimen CV. The primary audit uses UNI
\cite{chen2024uni} as the frozen encoder, and we repeat the identical protocol with
four further PFMs---Phikon, Phikon-v2, a Lunit DINO ViT, and a CTransPath-style
SimCLR ResNet---to test whether the conclusions are encoder-specific. All compute
runs in one HPC allocation via a resumable, filesystem-atomic worker pool.


\section{Results}\label{sec:results}
Unless stated otherwise, results use the frozen UNI encoder, the TabPFN probe,
pathway and gene targets, and specimen-level splits. We first establish that the
training-free audit recovers molecular signal that survives controls and rank encoders and tissues;
we then quantify label efficiency and in-context scaling, robustness under
distribution shift, and the negative and shortcut controls that separate trustworthy
decodability from apparent alignment.

\subsection{Decodability across encoders and tissues}
\begin{figure}[H]\centering
\begin{subfigure}{0.49\columnwidth}\includegraphics[width=\linewidth]{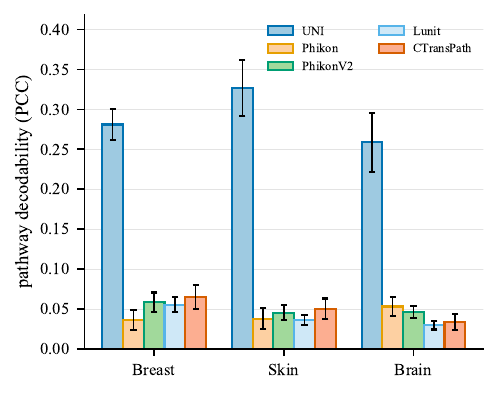}
\caption{Pathways}\label{fig:dec_path}\end{subfigure}\hfill
\begin{subfigure}{0.49\columnwidth}\includegraphics[width=\linewidth]{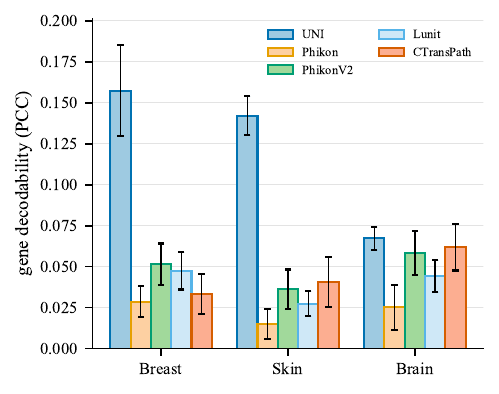}
\caption{Genes}\label{fig:dec_gene}\end{subfigure}
\caption{\textbf{Decodability across encoders.} Per-organ decodability (mean PCC, $K{=}512$;
$95\%$ CI over seeds) across five frozen pathology encoders (one modern large-scale PFM,
UNI, and four established pathology encoders) on Breast, Skin, and
Brain, for (a) pathways and (b) individual genes. Only the encoder changes, so the in-context
probe ranks representations directly (colourblind-safe Okabe--Ito palette).}
\label{fig:rq1}
\end{figure}
We first ask whether the training-free audit recovers molecular signal that survives
negative controls (Fig.~\ref{fig:rq1}): frozen pathology representations are paired
with spatially matched transcriptomic targets and decoded by TabPFN without
task-specific gradient updates. Across all five encoders, matched decodability sits
far above the label-shuffle and random-feature floors, so the recovered signal is a
property of the representation rather than of the probe or the target marginals. Two
structural patterns emerge. First, decodability is strongly \emph{tissue-dependent}:
structured, epithelial organs (Breast, Skin) show stronger signal than Brain in our
cohort. Second, \emph{pathway} activity is more consistently recoverable than
individual \emph{genes}, consistent with pathway aggregation reducing single-gene
noise. Both patterns are visible spatially in the per-spot reconstructions of
Fig.~\ref{fig:spatial}.

Holding the probe, targets, and splits fixed and swapping only the encoder turns the
audit into a standardized, gradient-free benchmark of representations. Pathway
decodability ranks UNI ($\ladTabPath$) far above the four substitutes---Phikon,
Phikon-v2, and CTransPath cluster near $0.054$ and Lunit trails at $0.044$---with UNI
significantly best in every organ$\times$target cell (Mann--Whitney over per-target
PCCs, Holm-corrected; all $p<10^{-5}$). UNI clears the label-shuffle floor by
$\sim$19$\times$ on pathways, whereas the substitutes sit only $\sim$3$\times$ above
it and within a hair of one another---weak in absolute terms. Encoder choice thus
dominates decodability under our common frozen probe. We emphasize that this margin is a
property of our \emph{frozen, training-free} probe rather than an absolute ranking of
pathology FMs: on HEST-1k's trained-probe leaderboard UNI, UNIv1.5, and H-Optimus-0
are near-tied and newer encoders can surpass UNI~\cite{jaume2024hest}. We therefore
claim UNI is strongest \emph{among the five encoders we test}, not the best PFM in
general; under gradient-trained heads the substitutes would likely recover part of
this gap.
\subsection{Label efficiency and in-context scaling}
\begin{figure}[H]\centering
\begin{subfigure}{0.48\columnwidth}\includegraphics[width=\linewidth]{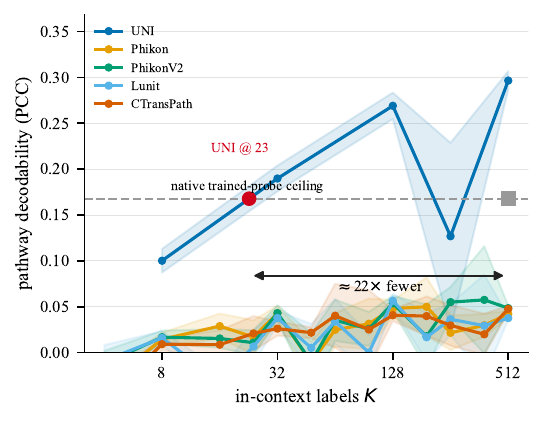}
\caption{Label efficiency}\label{fig:labeleff}\end{subfigure}\hfill
\begin{subfigure}{0.48\columnwidth}\includegraphics[width=\linewidth]{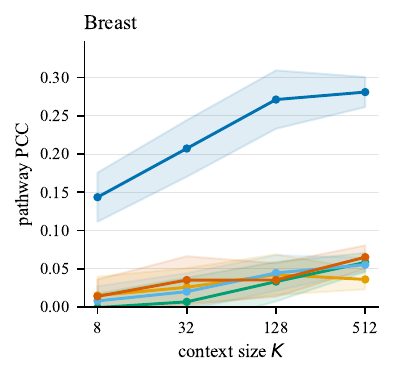}
\caption{Breast}\end{subfigure}\\[-0.5ex]
\begin{subfigure}{0.48\columnwidth}\includegraphics[width=\linewidth]{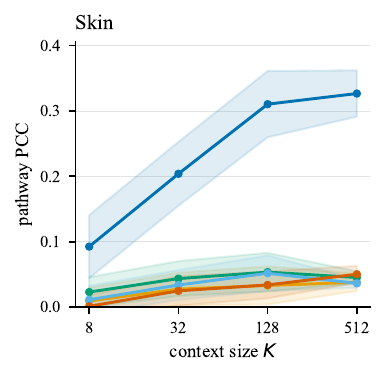}
\caption{Skin}\end{subfigure}\hfill
\begin{subfigure}{0.48\columnwidth}\includegraphics[width=\linewidth]{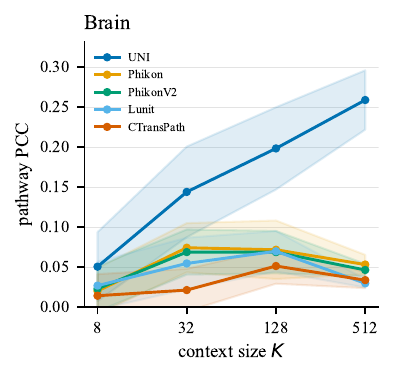}
\caption{Brain}\end{subfigure}
\caption{\textbf{Label efficiency and in-context scaling.} \textbf{(a)} UNI+TabPFN attains a
pathway PCC comparable to the reported HEST-1k trained-probe reference using
$\approx$23$\times$ fewer labeled context examples, while the other frozen encoders never reach
that level at any context size. \textbf{(b--d)} In-context scaling
per organ: pathway decodability versus context size $K$ for all five encoders ($95\%$ CI band over
seeds); only UNI rises with context.}
\label{fig:rq2}
\end{figure}
Fixing the encoder to UNI, we vary only the probe and the in-context budget
(Fig.~\ref{fig:rq2}). The training-free probes outperform the gradient-trained MLP:
TabPFN reaches pathway PCC $\ladTabPath$ versus $\ladMlpPath$ for the MLP
($p<10^{-3}$, Holm), and it attains a PCC comparable to the reported HEST-1k
trained-probe reference~\cite{jaume2024hest} using $\approx$23$\times$ fewer labeled
context examples---a level the other frozen encoders never reach at any context size.
This reference is a protocol-compatible point of comparison rather than a hard upper
bound, since the published trained probes differ in gene selection and preprocessing;
we therefore read it as a decodability reference, not a ceiling. For UNI, decodability rises with context size $K$,
while the substitute encoders stay flat. The same ranking is visible spatially
(Fig.~\ref{fig:spatial}): TabPFN's per-spot reconstruction tracks the
ground-truth tissue map more closely than kNN, Ridge, or MLP.
These absolute correlations (UNI pathway PCC $\ladTabPath$, gene PCC $\ladTabGene$)
are consistent with reported H\&E$\to$expression prediction on HEST-1k and related
spot-level benchmarks, which span roughly $0.14$--$0.43$ depending on gene selection
and protocol~\cite{jaume2024hest,he2020stnet,xie2023bleep}.
\subsection{Robustness under distribution shift}
\begin{figure}[H]\centering
\begin{subfigure}{0.32\columnwidth}\includegraphics[width=\linewidth]{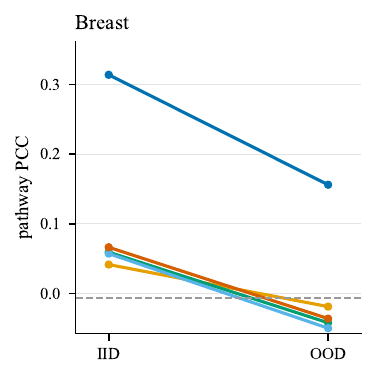}
\caption{Breast}\end{subfigure}\hfill
\begin{subfigure}{0.32\columnwidth}\includegraphics[width=\linewidth]{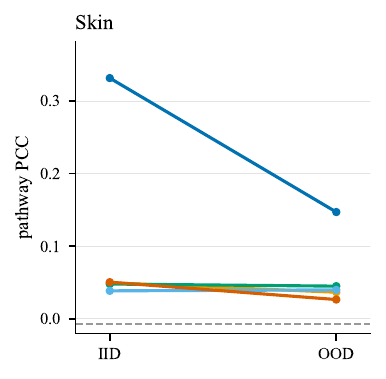}
\caption{Skin}\end{subfigure}\hfill
\begin{subfigure}{0.32\columnwidth}\includegraphics[width=\linewidth]{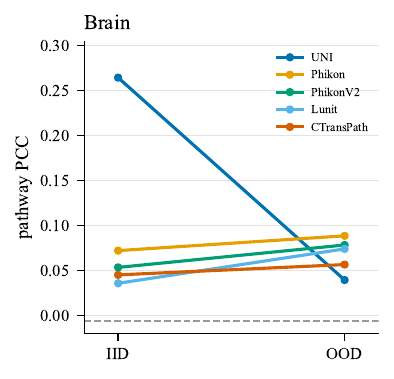}
\caption{Brain}\end{subfigure}
\caption{\textbf{Robustness under shift.} Per organ, the IID-to-organ-OOD drop in pathway
decodability for all five frozen encoders (dashed line = label-shuffle floor). High IID
scores do not guarantee transportable molecular alignment: UNI retains roughly half its
signal on Breast and Skin but collapses on Brain, while the substitutes have no signal to lose.}
\label{fig:rq3}
\end{figure}
We next stress-test the portability of the learned morphology--molecular mapping
across biological domains and image perturbations (Fig.~\ref{fig:rq3}). Organ holdout
is intentionally more demanding than section or cohort holdout and should not be
interpreted as an invariance test, because genuine molecular programs differ across
organs. In-distribution pathway PCC
$\oodIID$ falls to $\oodOrgan$ when the context is drawn from a held-out organ---a
$\oodDrop\%$ decline, with UNI retaining roughly half its signal on Breast and Skin
but collapsing on Brain, while the substitute encoders have no signal to lose. We
also re-embed queries under stain and colour jitter, rotation and flip, blur, and
JPEG compression across four severities, bootstrapping over the eight
species$\times$technology groups. UNI's molecular decodability is largely preserved:
geometric transforms (flip, $90^\circ$ rotation) are effectively lossless
(${\geq}\perturbRetainGeoMin\%$ of the clean $\perturbCleanPCC$ PCC), and even the
harshest photometric corruptions---heavy low-pass blur and strong colour
jitter---retain ${\approx}\perturbRetainMin\%$, with clean-vs-perturbed prediction
agreement staying above $\perturbAgreeMin$ throughout. For the one substitute we
re-embedded (Phikon-v2), the clean signal is smaller than the perturbation-induced
$\Delta\mathrm{PCC}$, so retention is undefined---there is no reliable signal to retain.
Finally, section identity is linearly decodable up to $214\times$ chance on human Visium,
exposing a strong site-specific shortcut documented in pathology
FMs~\cite{dejong2025medical,komen2026robust}. High in-distribution scores therefore
do not guarantee transportable molecular alignment: the trustworthiness axes must be
checked directly, which we do next.
\subsection{Trustworthiness controls}
\begin{figure}[t]\centering
\includegraphics[width=0.32\columnwidth]{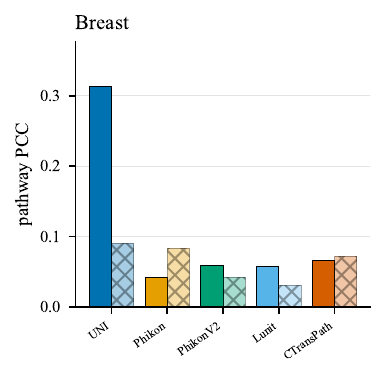}\hfill
\includegraphics[width=0.32\columnwidth]{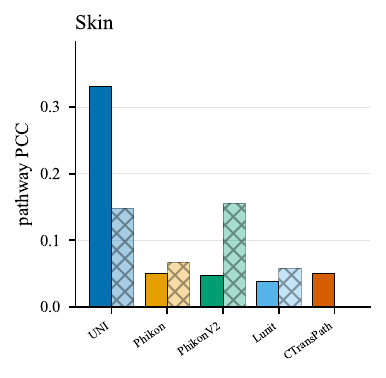}\hfill
\includegraphics[width=0.32\columnwidth]{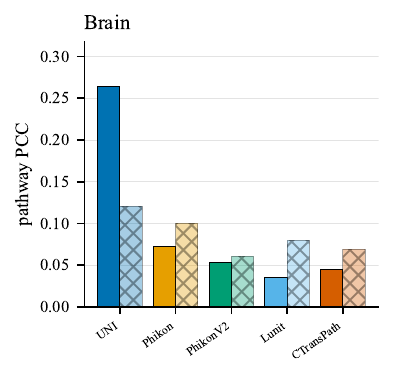}
\caption{\textbf{Trustworthiness controls.} Per organ, matched decodability ($K{=}512$, solid)
vs.\ the within-section control ($K{=}256$, the largest context feasible within one section, hatched) and the label-shuffle floor (dashed). UNI's
residual remains above the floor, although the unequal context sizes limit a direct effect-size comparison; the
substitutes' controls meet or exceed their matched scores.}
\label{fig:rq4}
\end{figure}
Finally, we test whether a high decodability score reflects shared molecular
information rather than a shortcut (Fig.~\ref{fig:rq4}). Matched decodability $\ctrlMatchPath$ collapses
toward chance under label permutation $\ctrlShufPath$ and random features
$\ctrlRandPath$, ruling out target-marginal and probe-only explanations, and
predictions remain stable across context reseeds. The within-section control is the
critical shortcut test: permuting targets \emph{within} a section removes the
cross-section shortcut. Because a single section supplies fewer spots, this control
uses the largest feasible context ($K{=}256$); under it UNI's pathway decodability
drops from $\ctrlMatchPath$ to $\ctrlWSecPath$ yet stays well above the
$\ctrlShufPath$ shuffle floor---a nonzero residual beyond the section shortcut (the
unequal context sizes, discussed below, preclude a clean effect-size estimate). For the
substitute encoders the control meets or exceeds the matched score, marking their
already-weak signal as section-driven. A clear matched--control
gap is therefore required before a readout is treated as trustworthy: high
decodability alone does not imply robust biological alignment.

\subsection{Qualitative spatial predictions}
\begin{figure}[H]\centering
\includegraphics[width=\columnwidth,trim=0 680 0 0,clip]{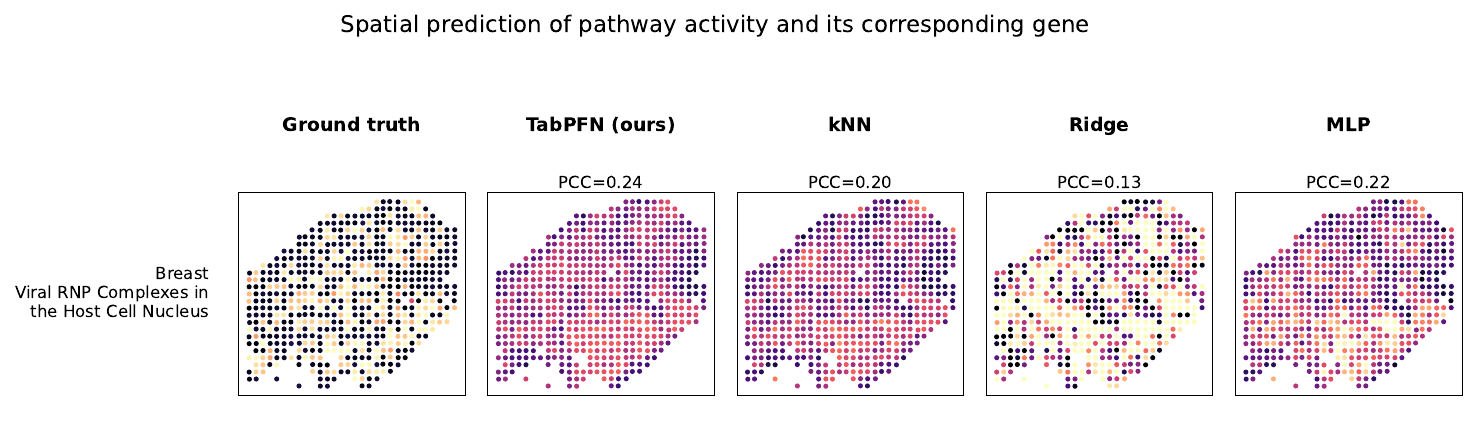}\\[-0.5ex]
\includegraphics[width=\columnwidth,trim=0 415 0 342,clip]{fig_spatial}\\[-0.5ex]
\includegraphics[width=\columnwidth,trim=0 140 0 610,clip]{fig_spatial}
\caption{\textbf{Spatial prediction of pathway activity.}
For one representative section per organ (Breast, Skin, Brain), rows show the
most-decodable Hallmark pathway. Columns show the ground-truth spatial map beside
reconstructions from the training-free TabPFN probe and the kNN, Ridge, and MLP
baselines, all sharing the \emph{same} frozen UNI features and $K{=}512$ context;
per-panel PCC is annotated. TabPFN reproduces the spatial pattern most faithfully,
and decodability is strongest in structured tissue (Skin) and weakest in
Brain---mirroring the quantitative trends in the main text.}
\label{fig:spatial}
\end{figure}

The aggregate correlations establish whether molecular programs are decodable, but
they do not show whether the predictions recover meaningful spatial organization
within tissue. We therefore complement the quantitative audit with qualitative
spatial reconstructions (Fig.~\ref{fig:spatial}). For one representative section from
each organ, we select that organ's most-decodable Hallmark pathway and reconstruct its
activity over the entire section with each probe. All methods use the same frozen UNI
features, specimen-level split, and $K{=}512$ context, making the panels directly
comparable with the quantitative results above. The spatial maps reinforce the main
findings: TabPFN most faithfully preserves the ground-truth organization, Skin shows
the clearest structured signal, and Brain remains comparatively difficult.
Figure~\ref{fig:spatial} thus provides a spatial counterpart to the PCC-based analysis,
showing that the strongest aggregate scores correspond to coherent tissue-level
patterns rather than isolated spot-wise agreement.

\section{Discussion}
Our results clarify how a frozen-probe score should be interpreted. A high PCC shows
molecular variation is accessible from a representation but does not identify its source:
the shuffled and random-feature controls rule out probe-only and target-marginal effects,
whereas within-section permutation asks whether local morphology carries information
beyond section membership. UNI retains a clear residual under this control, consistent
with morpho--molecular alignment that survives it, but the simultaneous organ-OOD drop
shows the alignment is not uniformly portable---reconciling strong in-distribution
decoding with weak deployment robustness when biological signal and acquisition-specific
structure coexist in the same embedding.

This also changes how pathology foundation models should be compared: a single trained
head or IID metric can reward optimization capacity or shortcuts shared by train and
test, whereas our frozen common probe holds optimization fixed and exposes which portion
of the score survives increasingly realistic stress tests. We therefore recommend
reporting a compact profile rather than one leaderboard number---matched decodability,
separation from negative controls, retention under shift, and the within-section
residual---which makes failure modes visible and distinguishes a merely predictive
encoder from one whose molecular readout supports biological interpretation. The pathway
advantage is practically useful here: because pathway scores aggregate coordinated
activity, pathway-level probing offers a more reproducible first screen for model
selection, with gene-level analysis reserved for programs that pass the shortcut controls.

\section{Limitations and Outlook}
Our conclusions are bounded by the probe, controls, and cohort. First, TabPFN is the
primary probe family. Ridge, kNN, and MLP baselines show that the broad tissue and
encoder trends are not unique to one head, but absolute PCC and especially label
efficiency can remain probe-specific. Replication with additional frozen tabular
models and matched-capacity trained probes is needed. Second, the within-section
control is limited to $K{=}256$, whereas the matched condition uses $K{=}512$.
Consequently, its lower PCC mixes shortcut removal with reduced context; we interpret
only the residual above the shuffle floor and leave a matched-$K$ ablation as necessary
follow-up.

Third, organ-OOD evaluates portability across distinct biological domains rather than
invariance of the underlying biology. Organ-specific programs, differing gene panels,
and technology composition may all contribute to the observed drop, even though our
primary verdicts use within-group contrasts and pathway aggregation reduces panel
sensitivity. Section- and cohort-OOD are therefore more direct robustness tests, while
organ-OOD should be read as a deliberately stringent transfer bound. The definitive
audit covers only Breast, Skin, and Brain because these organs contain enough sections
for the shortcut control; evaluating additional, lower-resource organs with uncertainty
intervals tailored to their smaller section counts is important for generality.

Finally, HEST lacks harmonized scanner, stain, and site metadata, preventing the most
deployment-relevant controlled shifts. Site identity is linearly decodable in PFM
embeddings in prior work~\cite{dejong2025medical,komen2026robust}, consistent with our
section probe, but acquisition-labeled external cohorts are required to quantify this
risk directly. Our label permutation controls also do not explicitly model spatial
autocorrelation; geostatistical baselines and Moran's-$I$-preserving nulls would better
separate morphology-anchored signal from spatial smoothness. Likewise, context-reseed
stability is a sensitivity diagnostic, not a formal uncertainty analysis: we do not
report ECE/TACE or predictive-coverage curves. Future work should combine calibrated
coverage diagnostics with spatially aware nulls, additional probe families, and
scanner/site-shift cohorts. Mechanistic SAE analyses~\cite{le2024sae} could further
identify which visual concepts support or confound the molecular programs that pass
our behavioral audit.

\section{Conclusion}
We developed a training-free framework for auditing whether frozen pathology
foundation models encode molecular state and whether that apparent alignment can be
trusted. By pairing frozen pathology embeddings with a frozen TabPFN probe, the audit
isolates information already present in the representation rather than information
learned by a task-specific prediction head. Across five encoders, three organs, gene
and pathway targets, and matched controls, four conclusions emerge. First, the decoded
signal survives negative controls: matched performance remains well above label-shuffled and
random-feature floors. Second, it is biologically patterned: pathway activity is more
consistently recoverable than individual-gene expression, and structured epithelial
tissues (Breast, Skin) are more decodable than Brain in our cohort. Third, strong
in-distribution scores do not guarantee transportability; UNI loses roughly
\oodDrop\% of its pathway signal under organ-level shift, and section identity is
linearly decodable at up to $214\times$ chance. Finally, the within-section control
retains a molecular residual above the shuffle floor, although unequal context sizes
prevent a clean estimate of shortcut magnitude. Together, these results show that decodability alone is not a
certificate of biological understanding. A pathology foundation model should instead
be evaluated jointly for molecular signal, negative-control separation,
distribution-shift robustness, and shortcut sensitivity. Because our probe is
standardized and gradient-free, this complete audit can serve as a practical routine
diagnostic for new pathology foundation models.

\bibliographystyle{aaai}
\bibliography{references}

\end{document}